\documentclass[pdflatex,sn-mathphys]{sn-jnl}

\usepackage{verbatim}
\usepackage{graphicx,subcaption}
\jyear{2023}%

\begin{document}

\title[Precise Image Generation on Noisy Quantum Computing Devices]  
        {Precise Image Generation on Current Noisy Quantum Computing Devices}  

\author*[1,2]{\fnm{Florian} \sur{Rehm}}\email{florian.matthias.rehm@cern.ch}
\author[1]{\fnm{Sofia} \sur{Vallecorsa}}\email{sofia.vallecorsa@cern.ch}
\author[2,3]{\fnm{Kerstin} \sur{Borras}}\email{kerstin.borras@desy.de}
\author[3]{\fnm{Dirk} \sur{Krücker}}\email{dirk.kruecker@desy.de}
\author[1]{\fnm{Michele} \sur{Grossi}}\email{michele.grossi@cern.ch}
\author[3]{\fnm{Valle} \sur{Varo}}\email{valle.varo@desy.de}

\affil[1]{\orgname{CERN}, \orgaddress{\city{Geneva}, \country{Switzerland}}}

\affil[2]{\orgname{RWTH Aachen University}, \orgaddress{\city{Aachen}, \country{Germany}}}

\affil[3]{\orgname{DESY}, \orgaddress{\city{Hamburg},  \country{Germany}}}

\abstract{
The Quantum Angle Generator (QAG) is a new full Quantum Machine Learning model designed to generate accurate images on current Noise Intermediate Scale (NISQ) Quantum devices. Variational quantum circuits form the core of the QAG model, and various circuit architectures are evaluated. In combination with the so-called MERA-upsampling architecture, the QAG model achieves excellent results, which are analyzed and evaluated in detail. To our knowledge, this is the first time that a quantum model has achieved such accurate results. To explore the robustness of the model to noise, an extensive quantum noise study is performed. In this paper, it is demonstrated that the model trained on a physical quantum device learns the noise characteristics of the hardware and generates outstanding results. It is verified that even a quantum hardware machine calibration change during training of up to 8\% can be well tolerated. For demonstration, the model is employed in indispensable simulations in high energy physics required to measure particle energies and, ultimately, to discover unknown particles at the Large Hadron Collider at CERN.
}


\maketitle
\section{Introduction}\label{sec1}
Quantum computing has the potential for a new paradigm in future computing to accelerate tasks or even handle classically unsolvable problems \cite{unsolvable_problems}. In the current Noise Intermediate Scale Quantum (NISQ) era, quantum devices suffer from non-negligible hardware errors, limited connectivity and a limited number of qubits \cite{NISQ}. While practical quantum advantage is currently extremely difficult to accomplish, finding the best suited algorithms to effectively combat the problems of NISQ-era devices remains a widely researched topic. Quantum Machine Learning (QML) is a domain which achieves acceptable results on NISQ devices due to the observed robustness against noise \cite{qml_data}. 

High Energy Physics (HEP) experiments, such as those at the Large Hadron Collider (LHC) at CERN, require enormous amounts of simulated data for deriving high precision physics results \cite{bigdata_2021}. To handle this demand, gigantic quantities of computing hardware resources are necessary, which has led to the creation of the world's largest computing grid operated by CERN \cite{WLCG_webpage}. To alleviate this strain on computational resources, Machine Learning (ML) models have been developed that exhibit remarkable speed-ups over current Monte Carlo-based simulations while maintaining the required level of accuracy. In general, QML simulations represent a promising approach to address the only further increasing simulation demands in the future \cite{QML_speedup,qml_hep}. QML employs quantum circuits which exploit the quantum properties of superposition and entanglement, which possess the potential to outperform neural networks, their classical analogue \cite{quantum_supremacy}. In addition, QML might have the advantage of learning more complex distributions with fewer parameters than classical ML due to their wider accessible phase space.

Encoding the classical data into qubit states on quantum computers is a non-trivial task \cite{state_preparation2}. Currently, many encoding techniques exist, each exhibiting specific advantages and disadvantages, and in practice, identifying “the best” encoding technique remains application dependent \cite{state_preparation2}. To achieve a potential quantum advantage over classical computing, theoretical studies suggest that at least linear scaling from qubits to features is required \cite{data_encoding}. On the other side, models which employ better than linear scaling encoding techniques have drawbacks, making them unsuitable for generating precise images on NISQ devices. For example, amplitude encoding can only generate probability distributions and not absolute pixel entries, i.e. energy values.  

At present, there exist several quantum generative models, for example, the Quantum Circuit Born Machine (QCBM) \cite{born_machine}, Quantum Variational Autoencoders \cite{Qvariational_autoencoder} or variations of quantum Generative Adversarial Networks \cite{qgan_survey, 1QGAN_hep,QuGAN}. They all face limitations. Some models either do not scale well in terms of qubits required relative to the number of encoded features, or they do not achieve a satisfying level of fidelity. The Quantum Angle Generator (QAG) presented in this paper and first introduced in reference \cite{ACM_IEEE_QAG}, aims to overcome these problems. Employing angle encoding with linear scaling of qubits to features, it achieves extremely accurate results for a real-world problem on current physical noisy quantum devices. 

The paper content is structured as follows. First, the HEP use case is motivated and the training data is defined. Next, the QAG model and the employed angle encoding technique are presented. Then, multiple circuit architectures are compared and the advantages of the best ones are highlighted. An in-depth accuracy analysis of the model follows to highlight its excellent precision. The quantum hardware noise behavior is evaluated, including training and inference executed on real quantum devices. Lastly, conclusions are drawn and summarized.

\section{High Energy Physics Simulations}
Simulations remain a crucial component of HEP analysis to evaluate the results obtained by the processed data of the experiments. Currently, simulations are performed predominantly with Monte Carlo methods such as the Geant4 toolkit \cite{Geant4}. However, Monte Carlo simulations are very hardware resource demanding and occupy half of the worldwide LHC Computing Grid \cite{RoadmapHEP}. Future LHC experiments will require more simulations due to more energetic particles, simultaneous collisions and detectors constructed with higher granularity. However, the projected budget for hardware development and computing resources cannot keep pace with these increasing demands \cite{detector_simulations, LHC_computing_demand}. As a result, ML alternatives to Monte Carlo methods are being actively researched. Initial prototypes predict significant reductions in simulation time and hardware resources while retaining acceptable levels of accuracy \cite{ju2021performance, biscarat2021towards, sela2022deep}. This research goes one step beyond classical ML. Since HEP data sets are generally created by underlying quantum mechanical effects, performing the computations on quantum devices which likewise make use of quantum effects has the potential to substantially enhance the simulations in accuracy and in terms of sustainable computing. 

\begin{figure}[t!]  
    \centering
    \includegraphics[width=0.7\textwidth, clip=true]{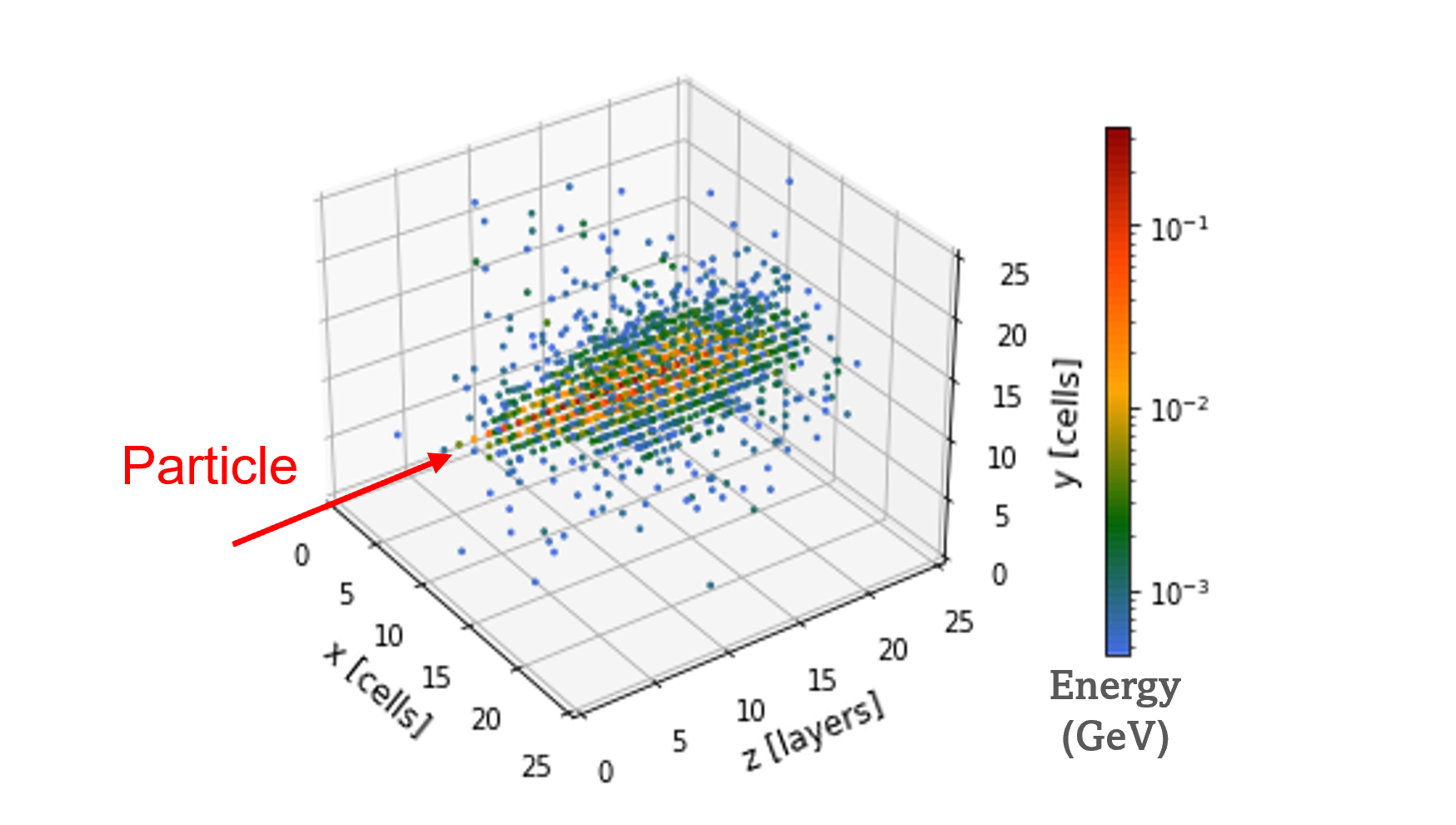}
    \caption{An example 3D calorimeter shower image. A primary particle enters the calorimeter at $(x, y, z)$ = $(13, 13, 0)$ and generates a secondary particle shower along the z-axis.}
    \label{fig:training_data}
\end{figure}

Electromagnetic calorimeters are constructed as high granularity sensor grids to measure the energy of photons, electrons, and positrons through complex particle shower generation processes in space and time \cite{calorimeter_cern}. They constitute a key component of HEP detectors to measure the energy of the particles produced in the interaction process and occupy most of the simulation time \cite{RoadmapHEP}. Calorimeter outputs can be interpreted at lowest order as static and spatial 3D images, which we call “shower images”: the value of each pixel corresponds to the energy measured in a specific calorimeter cell. The initial data from reference \cite{data_set} consists of $25 \times 25 \times 25$ pixel images. An example of a 3D shower image is visualized in figure \ref{fig:training_data}. To reduce the dimensionality, the images are averaged along two spatial axes (x- and y-direction), resulting in a one-dimensional representation that is further downsampled to eight pixels by averaging three contiguous pixels along the z-direction. Although the initial data set provides many different energies, for simplicity this study focuses on images recorded by particles in the energy range of [225, 275]\,GeV. The data set is split into a training and test set, each consisting of approximately 1\,000 samples. The downsampled data is available in reference \cite{quantum_data_set}, and an example image is illustrated later in this paper in figure \ref{fig:showershape}.

\section{The Quantum Angle Generator}
The QAG represents a QML model which employs the well established technique of angle encoding \cite{encoding_classifiers, angle_encoding_application} to generate extremely precise images. 
It scales linearly with the number of encoded features. Thus, the generation of $n$ features requires $n$ qubits. In this study, the number of features corresponds to the number of pixels. A comprehensive description of the QAG model, its objective training function, and an evaluation of several quantum circuits are provided below. 

\begin{figure}[t!]  
    \centering
    \begin{subfigure}[t]{.8\textwidth}
        \includegraphics[width=\textwidth]{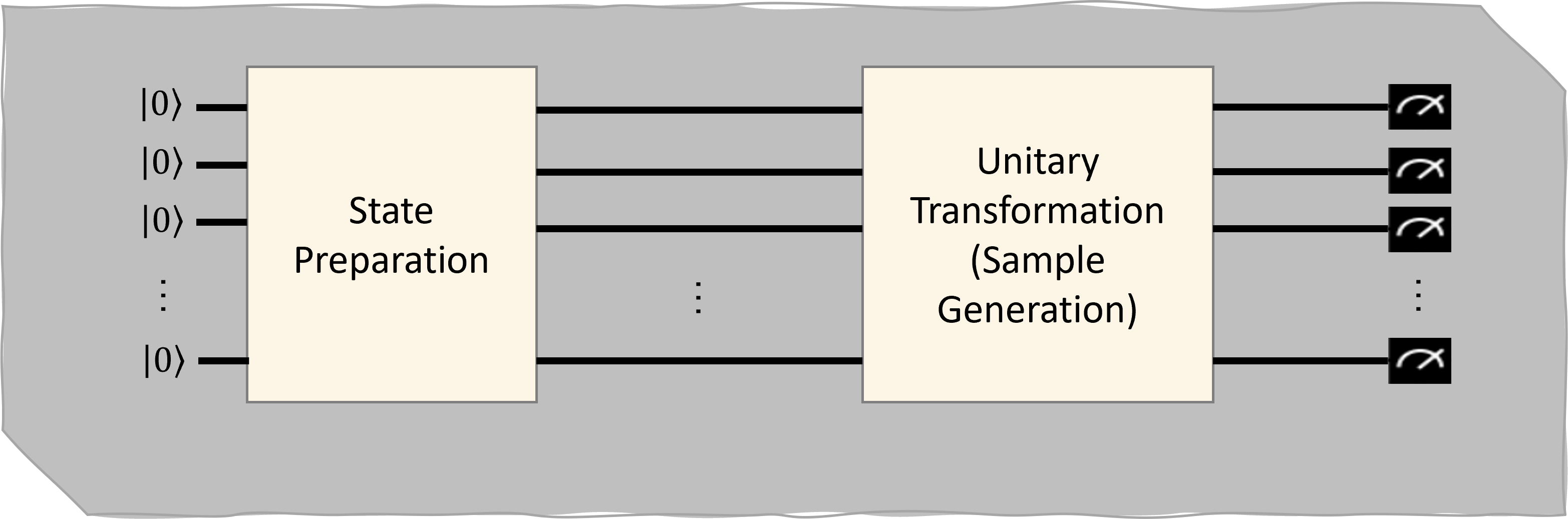}
    \end{subfigure}
    \caption{The structure of the Quantum Angle Generator.}
    \label{fig:model_structure}
\end{figure}

\subsection{Model Description}
\label{sec:QAG_description}
The QAG model consists of variational quantum circuits trained by an objective function. The model structure is visualized in figure \ref{fig:model_structure}. All qubits are initialized in the basis state $\vert 0 \rangle$. The state preparation function implements a Hadamard (H) gate to constitute superposition, followed by a y-rotational (Ry) gate to introduce randomness so that the model can draw new samples at each execution. For this, the Ry gate angles $\Omega$ are randomly drawn from a [-1, 1] uniform distribution and pixel-wise multiplied by the pixel standard deviations present in the training data to obtain correct pixel energy variations. To account for all various primary particle energies, all angles $\Omega$ are multiplied by a random value between [-0.25, 0.25].

The unitary transformation consists of quantum circuits and constitutes the trainable part of the QAG model. Various circuit architectures were tested as documented in section \ref{sec:circuits}.

To convert the quantum states back into classical energy values via angle encoding, the model must be executed multiple times and the quantum states measured. The number of executions is commonly denoted as the number of shots $nb_{\textrm{shots}}$. It is counted how often state $\vert 0 \rangle$ is measured. The scalar intersection $I$ of the vertical axis on the Bloch sphere (z-axis) and the angle $\theta$ is calculated with:
\begin{equation}
\begin{aligned}\label{eq:Intersection}
    I &= 2 \cdot \frac{\textrm{counts}(\vert 0 \rangle)}{nb_{\textrm{shots}}}-1\,,\\[0.5em] 
    \theta &= \arcsin (I)\,.
\end{aligned}
\end{equation}

\begin{figure}[t!]  
    \centering
        \begin{subfigure}[t]{.37\textwidth}
        \subcaption{}
        \includegraphics[width=\textwidth]{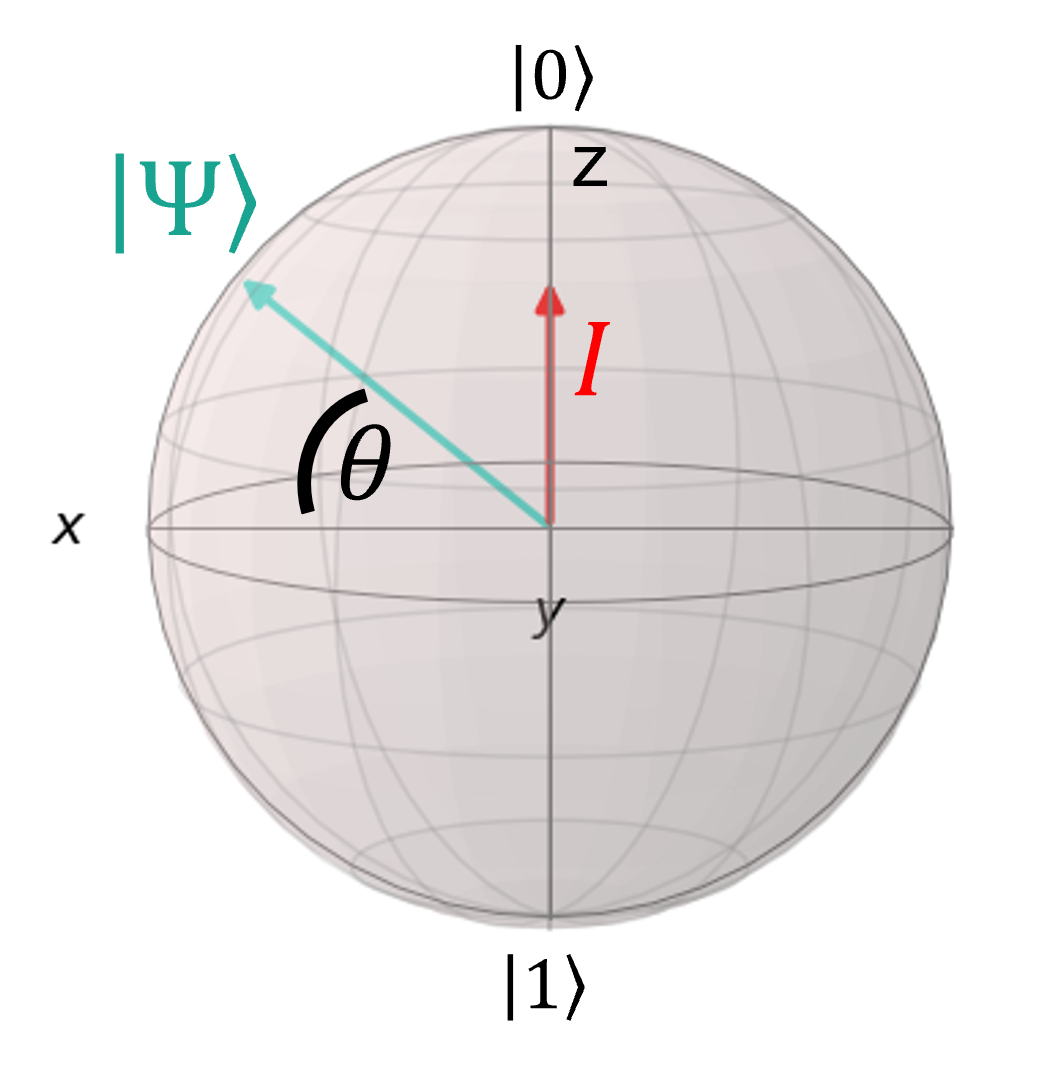}
        \label{fig:encoding_a}
    \end{subfigure}
    \hfill
    \begin{subfigure}[t]{.54\textwidth}
        \subcaption{}
        \includegraphics[width=\textwidth]{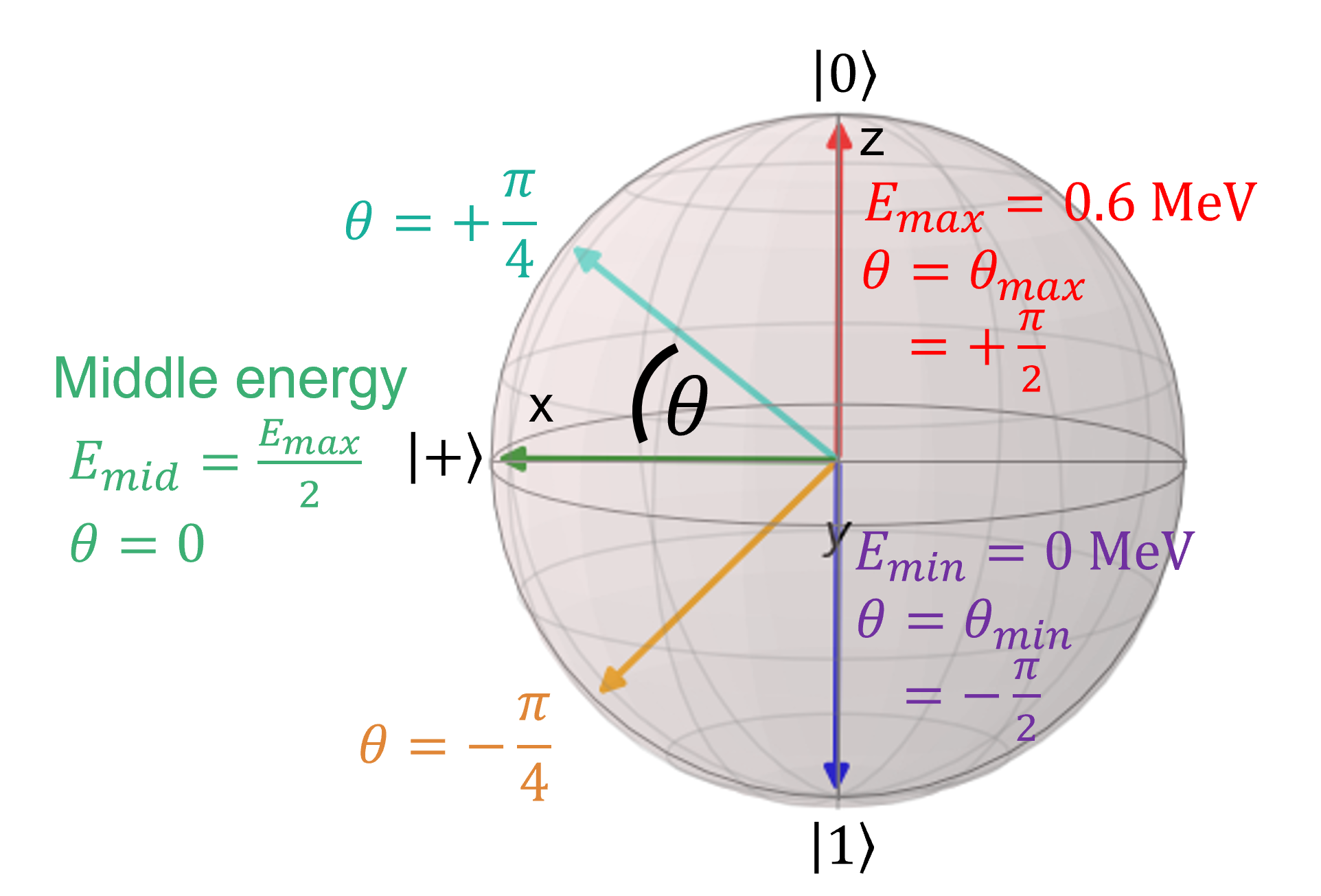}
        \label{fig:encoding_b}
    \end{subfigure}
    \caption{(a) Decoding of an example state $\Psi$ into an angle $\theta$. (b) Some representative angles $\theta$ and its corresponding energy values. }
\end{figure}

The angle $\theta$ operates in the x-z-plane of the Bloch sphere and is defined as zero in the $\vert + \rangle$ state. Rotating $\theta$ clockwise leads to positive angles. The decoding process is visually illustrated in figure \ref{fig:encoding_a} with an example state $\vert \Psi \rangle$, its intersection $I$ and the corresponding angle $\theta$ for a single qubit example. The angle $\theta$ can then be transformed into a pixel energy $E$ by the linear change of ranges equation:
\begin{equation}
    E = \frac{(E_{\textrm{max}}-E_{\textrm{min}})(\theta - \theta_{\textrm{min}})}{\theta_{\textrm{max}}-\theta_{\textrm{min}}}   \,.
    \label{eq:decoding}
\end{equation}
$E_{\textrm{max}}$, $E_{\textrm{min}}$, $\theta_{\textrm{max}}$ and $\theta_{\textrm{min}}$ are defined in figure \ref{fig:encoding_b}: the minimum energy $E_{\textrm{min}} = 0\,$MeV is set to $\theta = -\pi/2$ and the maximum energy $E_{\textrm{max}} = 0.6\,$MeV is set to $\theta = +\pi/2$. 
With $E_{\textrm{min}}=0$ and $\theta_{\textrm{max}}= -\theta_{\textrm{min}}$ equation \ref{eq:decoding} can be simplified into:
\begin{equation}
     E = \left( {\frac{E_{\textrm{max}}}{2 \cdot \theta_{\textrm{max}}}}\right) \cdot (\theta + \theta_{\textrm{max}})  \,.
    \label{eq:decoding2}
\end{equation}
For multi-qubit quantum circuits: the final quantum states of all qubits are measured, and the outcomes are decoded independently. Although the qubit results are individually decoded, the state of each qubit is entangled with others due to the gates applied within the variational quantum circuit, as lined out later.

It is worth to be noted that the angle $\theta$ and, therefore, the corresponding decoded energy $E$ remain in discrete values. Since, $\theta$ depends in value and accuracy on the number of shots $nb_{\textrm{shots}}$: the larger $nb_{\textrm{shots}}$, the better the achievable energy precision and resolution. Fortunately, with present quantum devices, the $nb_{\textrm{shots}}$ can be easily chosen to be large. Currently, on IBMQ devices the maximum possible number of shots is $nb_{\textrm{shots}} =$ 100\,000. For the simplified calorimeter use case, this is more than sufficient. For comparison, in a previous classical reduced precision ML research in reference \cite{reduced_precision}, it is demonstrated that already $256$ discrete energy levels are sufficient for correctly reproducing the full-size calorimeter shower image. In this reduced precision research, the parameters of the neural network are quantized from a larger format (floating point 32) down to a smaller number format (integer 8). This study will show that 512 shots provide a sufficient resolution. The detailed image generation process for the QAG model, as described in this subsection, is provided as an algorithm in the appendix \ref{app:QAG_algorithm}.

\subsection{Training Objective Function}
The QAG model is trained with two losses employed as objective functions. The first one is the Mean Maximum Discrepancy (MMD) loss \cite{MMD1,MMD2}, already successfully applied by other quantum models, e.g. the QCBM \cite{born_machine}. Training exclusively with the MMD loss resulted in good average shower distributions. However, when exploring the generated images in more detail, for example in the pixel correlation, the model did not perform satisfactorily. Therefore, a second correlation (Corr) loss is added to help learn the patterns present in the training data (e.g., image pixel correlations). The Corr loss is calculated by the pixel-wise mean squared error (MSE) between the pixel correlation values present in the training data and the ones inside the generated data. The pixel-wise correlations are illustrated in figure \ref{fig:best_corr_geant4}.

To train the QAG model, the Simultaneous Perturbation Stochastic Approximation (SPSA) optimizer \cite{SPSA} is employed, which only requires two optimization steps per epoch. The hyperparameters for training were found by extensive hyperparameter searches employing the Optuna \cite{optuna} library. All tests in this study are executed in Qiskit version 0.26.2. The models are trained for 500 epochs, containing one batch. The dynamic MMD loss weight starts at a value of one and decays linearly with $-0.001 \cdot $epoch, starting from epoch 100. Opposite, the Corr loss weight increases by the same value starting at zero. The dynamical training batch size is set to generate one image in the first hundred epochs and afterward to 20 images to calculate the Corr loss between multiple images. Each quantum job contains 512 shots for training and inference. The generator SPSA optimizer learning rate is set to $c_0=1$ with an exponential learning rate decay of $0.006$ starting from epoch 50. All these settings showed the best performance in the tests.

\subsection{Quantum Circuit Study}
\label{sec:circuits}
The ideal circuit should contain a certain, optimized to be minimal, number of parameters to achieve a sufficient level of accuracy. Different circuit architectures were compared to each other based on characteristic numbers expressing the power of the circuit. The circuits employ trainable rotational gates and two-qubit entanglement gates. As the angle encoding primarily uses the qubits y-axis component, we predominantly employ Ry gates. For some circuit architectures, we test if additional z-rotational gates (Rz gates) or deeper circuits with depth 2 (denoted as d2) can further improve the results. We use two-qubit controlled-not gates (cx gates) native on IBM Quantum (IBMQ) \cite{ibmq} devices; while other entanglement gates are compositions of multiple native gates of the hardware. Keeping an eye on the goal of executing the training on a real quantum device, the absolute number of decomposed gates should be kept as small as possible. 

The characteristic circuit numbers used in this study are the number of trainable parameters $N_p$, the expressibility $X$ and the entanglement capability $E$. Larger circuits with more trainable parameters are potentially capable of achieving more accurate results. However, it might be that a plateau is reached at some point where the same task can be solved with similar accuracy by a smaller circuit, which is being investigated here. The definitions for $X$ and $E$ are from reference \cite{expressibility}: $X$ describes how well the circuit can represent the pure states of the representative Hilbert space. For a single qubit the expressibility exhibits how many states of the Bloch sphere can be represented. In this paper, we measure $1-X$: the closer to $1$ the better the expressibility of the model, while the closer to $0$ the worse. The entanglement capability $E$ is a measure that expresses the ability of a circuit to generate entangled states between the qubits. Likewise, $E$ ranges from $0$ to $1$, where $1$ represents the best achievable value. The circuit architectures under study are introduced in the appendix \ref{sec:A1}. Their corresponding characteristic numbers and theoretical potential are provided in the appendix \ref{sec:A2}. In the following, the circuits are evaluated for the calorimeter use case.

\begin{figure}[t!]  
    \centering
    \includegraphics[width=0.99\textwidth, clip=true]{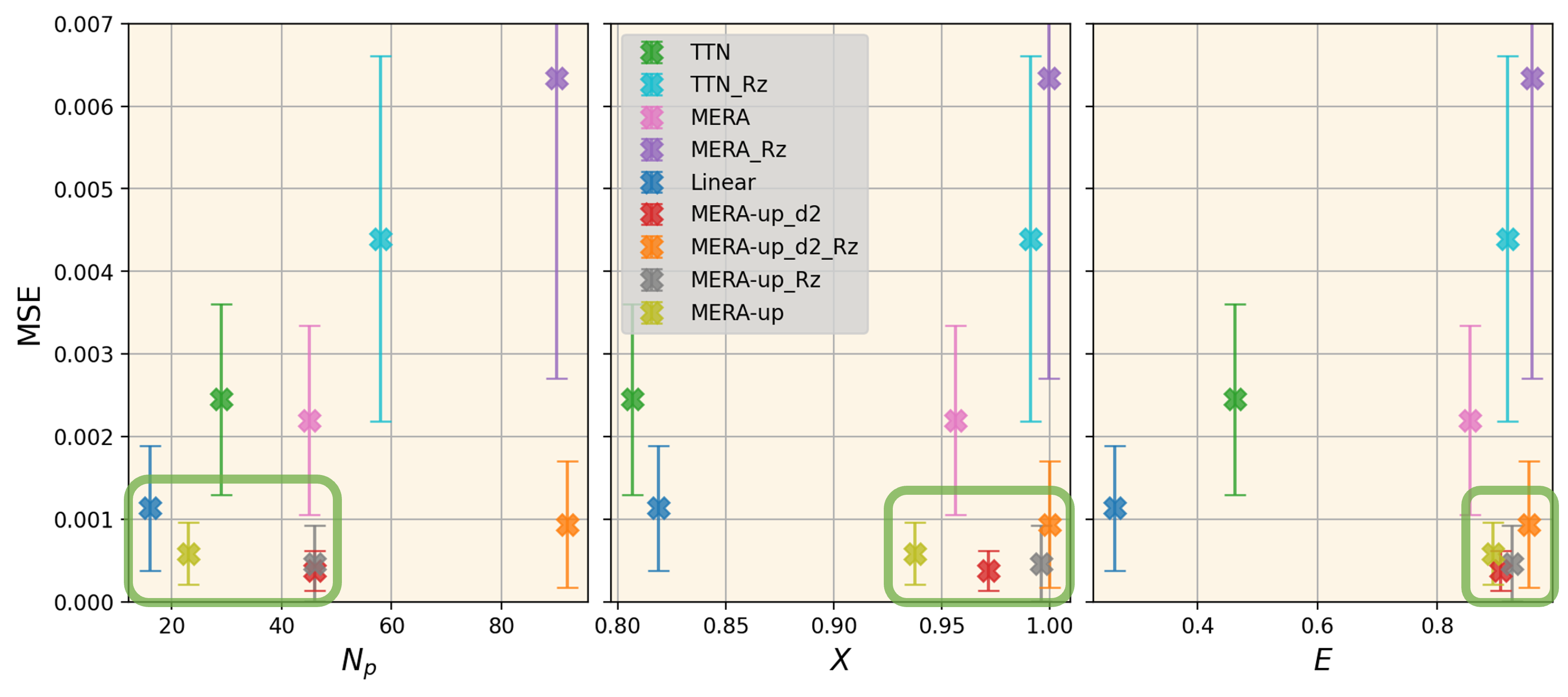}
    \caption{The MSE uncertainty (the smaller, the better) of the various trained circuit architectures as a function of $N_p$ (left), $X$ (middle) and $E$ (right), see text for explanations. Green rectangles mark optimal areas.}
    \label{fig:circuits_MSE}
\end{figure}

We start by interpreting the results displayed in figure \ref{fig:circuits_MSE}. The MSE accuracy metric is calculated by taking the pixel-wise MSE between the average Geant4 and QAG images. The training is repeated $25$ times for each circuit and the mean and standard deviations are plotted. To prevent the influence of outliers, the best and worst two trials are discarded in this analysis. The MSE is given as a function of: $N_p$ on the left, $X$ in the middle, and $E$ on the right. By inspecting the plots, it can be recognized that the MSE does not correlate with any of the characteristic circuit values in the plots, neither do the characteristic values correlate among themselves, as shown in the appendix \ref{sec:A2}. 

The MERA-up, MERA-up\_d2, and MERA-up\_Rz architecture perform best with the lowest MSE. This is consistent with the observation that they maintain a high $X$ and $E$, as provided in the appendix \ref{sec:A1}. The error bars provide a hint about the training stability. It can be observed that the better the average MSE of a model, the smaller its standard deviation.

All in all, the MERA-up\_Rz circuit clearly performs best considering the characteristic circuit values and the achieved accuracy in training. However, with the emphasis on a low number of $N_p$, the plain MERA-up circuit performs almost as well, while needing only half the number of parameters. Therefore, the following studies employ the MERA-up circuit for training the QAG model.

\section{In-depth Accuracy Analysis}
In this section, we analyze the results of the QAG model operating the MERA-up circuit architecture. We showcase typical accuracy metrics for the calorimeter simulation in HEP.

\begin{figure}[t!]  
    \centering
    \begin{subfigure}[t]{.75\textwidth}
        \subcaption{}
        \includegraphics[width=\textwidth]{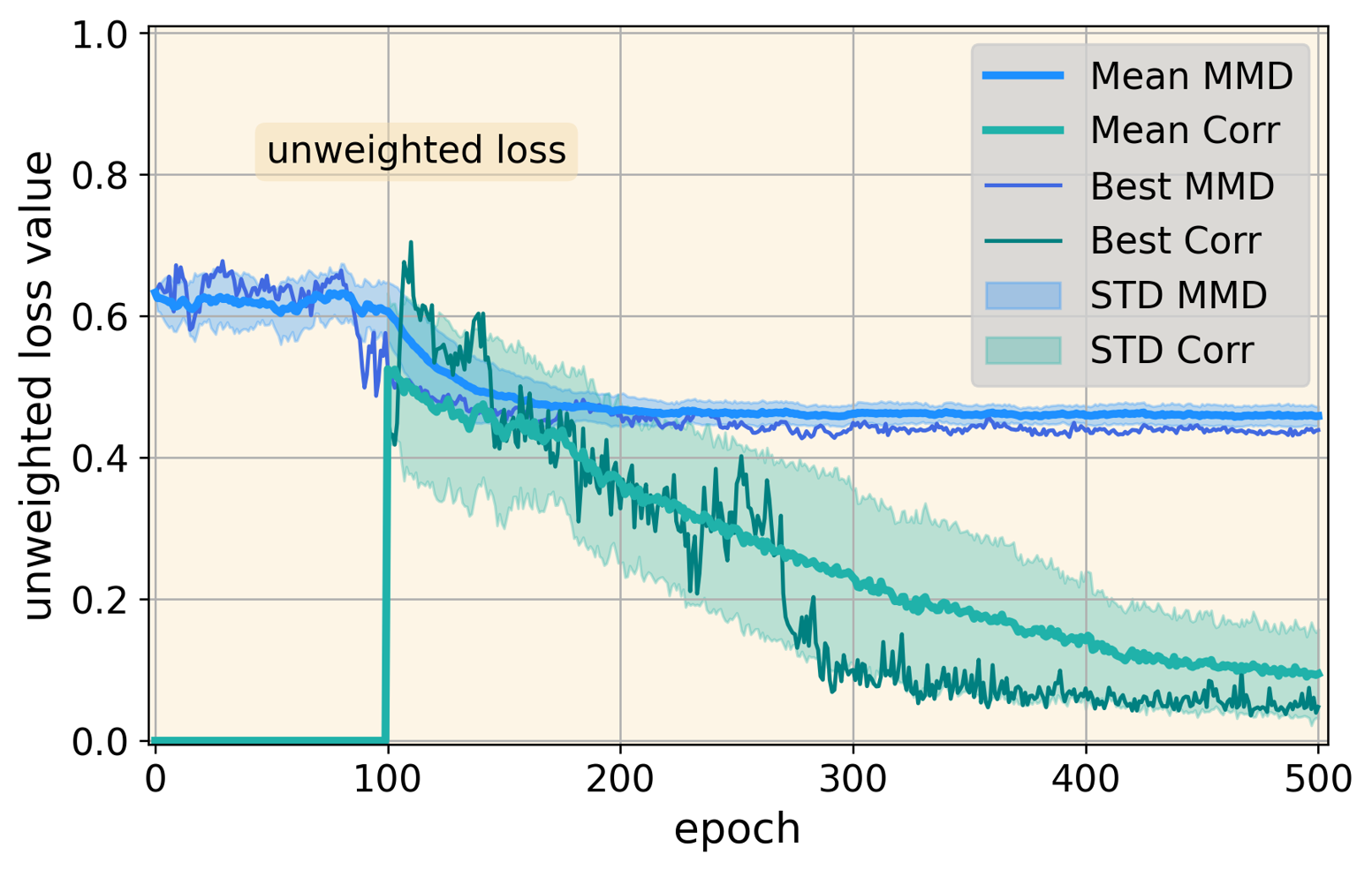}
    \end{subfigure}
    \caption{The unweighted loss values the training epochs. The blue lines correspond to the MMD losses and the green lines to the Corr losses. The thick solid line represents the mean loss across trials, the colored band around indicates the standard deviation (STD), and the thin line denotes the loss of the best trial's loss.}
    \label{fig:best_unweighted}
\end{figure}

\subsection{Training Evaluation}
In a first step, the statistical trends of the objective functions during training are investigated. In figure \ref{fig:best_unweighted}, the unweighted loss functions (excluding loss weights) are plotted as a function of the training epochs. The mean of twenty training repetitions is visualized as a thick solid line and the standard deviation (STD) as a colored band. The Corr loss starts influencing the training only at epoch 100 because its weight is set to zero before. 

The MMD loss of all training repetitions converges stable, and the STD band narrows towards the end of the training. Overall, the MMD and Corr loss converge smoothly without strong oscillations, which is a desirable characteristic for stable (Q)ML training. The MMD loss contributes far more than the Corr loss throughout the training. However, the Corr loss plays a significant role in achieving good physics accuracy in the generated shower images.

\subsection{Inference Evaluation}
\label{sec:accuracy_evaluation}
In the following, the accuracy in inference is evaluated. The generated images of the best trained model are compared to the Geant4 test data, which consists of 980 images. Likewise, there are 980 images generated by the QAG model to create the following accuracy metrics. The details about how the accuracy metrics are calculated are provided in the appendix \ref{app:validation_code}.

\begin{figure}[t!]  
    \begin{subfigure}[t]{.25\textwidth}
        \subcaption{}
        \includegraphics[height=\textwidth]{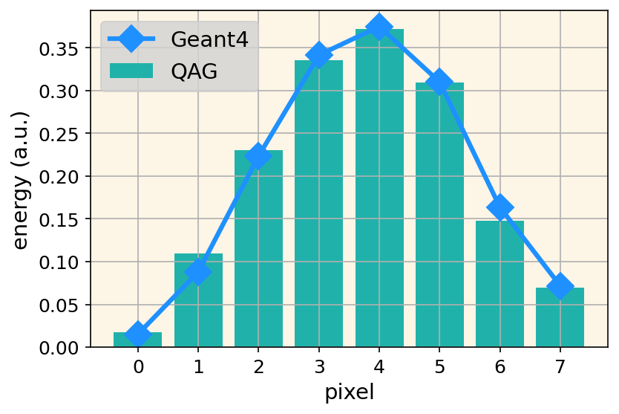}
        \label{fig:showershape}
    \end{subfigure}
    \hspace{1.25cm}
    \begin{subfigure}[t]{.255\textwidth}
        \subcaption{}
        \includegraphics[height=\textwidth]{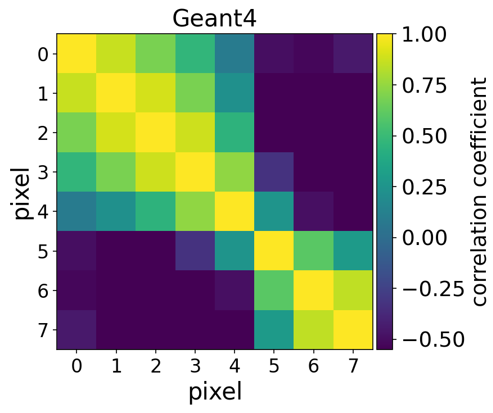}
        \label{fig:best_corr_geant4}
    \end{subfigure}
    \hspace{0.4cm}
    \begin{subfigure}[t]{.255\textwidth}
        \subcaption{}
        \includegraphics[height=\textwidth]{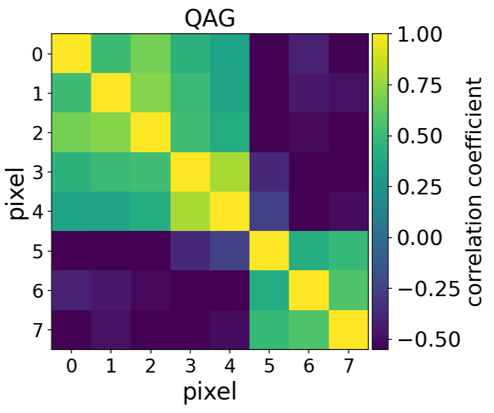}
        \label{fig:best_corr_qag}
    \end{subfigure}
    \caption{(a) Visualization of the average calorimeter shower shapes. The energy is given in an arbitrary unit (a.u.) due to image downsampling. The pixel-wise correlation plot for (b) Geant4 and (c) the QAG model. The correlation ranges between -1 and 1; a value of 1 indicates a perfect positive correlation.}
\end{figure}

\textbf{1. Average calorimeter shower shape}: 
The first metric represents the calorimeter shower shape displayed in figure \ref{fig:showershape}. The shower shape is perfectly reproduced by the QAG model. The MSE corresponds to $0.00059\pm0.00037$, which is extremely close to zero, indicating a very good accuracy.

\textbf{2. Pixel-wise correlation}: 
The second metric corresponds to the pixel-wise image correlation. The positive or negative correlation patterns between all the pixels are determined. The baseline represents the correlation from Geant4 in figure \ref{fig:best_corr_geant4}. The correlation for the generated data of the QAG model is presented in figure \ref{fig:best_corr_qag}. It can be derived that the overall correlation pattern is accurately reproduced by the QAG model. Like the Geant4 data, it consists of a larger and more compact positively correlated group of pixels. The other pixels are negatively correlated. Inspecting the particular details, different color shades indicate some minor deviations. However, the achieved correlation precision by QAG is astonishing. Therefore, it can be concluded, that the quantum circuits are capable to reproduce complex correlation patterns through substantial entanglement strategies, as present in the MERA-up architecture. 

\begin{figure}[t!]  
    \begin{subfigure}[c]{.25\textwidth}
        \subcaption{}
        \includegraphics[height=\textwidth]{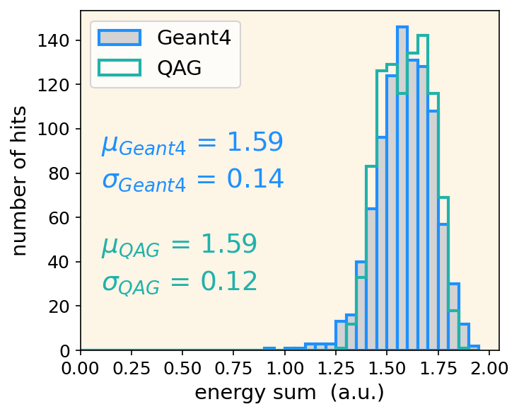}
        \label{fig:best_energy_sum}
    \end{subfigure}
    \hspace{0.55cm}
    \begin{subfigure}[c]{.28\textwidth}
        \subcaption{}
        \includegraphics[height=\textwidth]{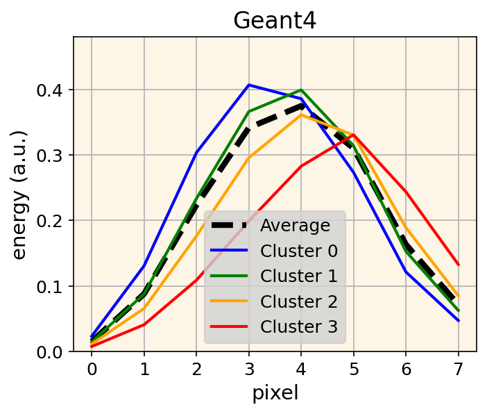}
        \label{fig:best_clusters_geant4}
    \end{subfigure}
    \hspace{0.4cm}
    \begin{subfigure}[c]{.28\textwidth}
        \subcaption{}
        \includegraphics[height=\textwidth]{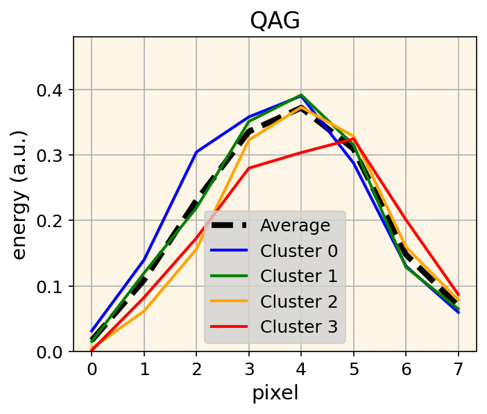}
        \label{fig:best_clusters_qag}
    \end{subfigure}
    \caption{(a) The energy sum histogram, the four k-means clusters present in the images of (b) the Geant4, and (c) the QAG images.}
\end{figure}

\textbf{3. Energy sum}: 
The energies contained in all pixels calculated for the individual images represent the third accuracy metric. In figure \ref{fig:best_energy_sum}, the energy sum histogram of the Geant4 images reveals a Gaussian shape and is correctly reproduced visually by the QAG model, which is also confirmed by the mean $\mu$ and the standard deviation $\sigma$.

\textbf{4. k-means Clusters}: 
This metric evaluates if the QAG model can correctly represent specific image modes. The Geant4 data is clustered with the k-means algorithm \cite{k_means} to find four clusters or image modes as illustrated in figure \ref{fig:best_clusters_geant4}. Cluster 0 deposits substantially larger fractions of energy in earlier calorimeter cells than in higher cluster numbers. Further, the particles from clusters 0 and 1 contain a larger energy fraction, estimated by the integral area below the curves. Here we are interested in whether the QAG model can reproduce this behavior. The four clusters of the QAG images are provided in figure \ref{fig:best_clusters_qag}. A similar structure can be observed, which indicates a good accuracy in reproducing the energy contents and image modes on average. 

\begin{figure}[t!]  
    \centering
    \includegraphics[width=0.95\textwidth, clip=true]{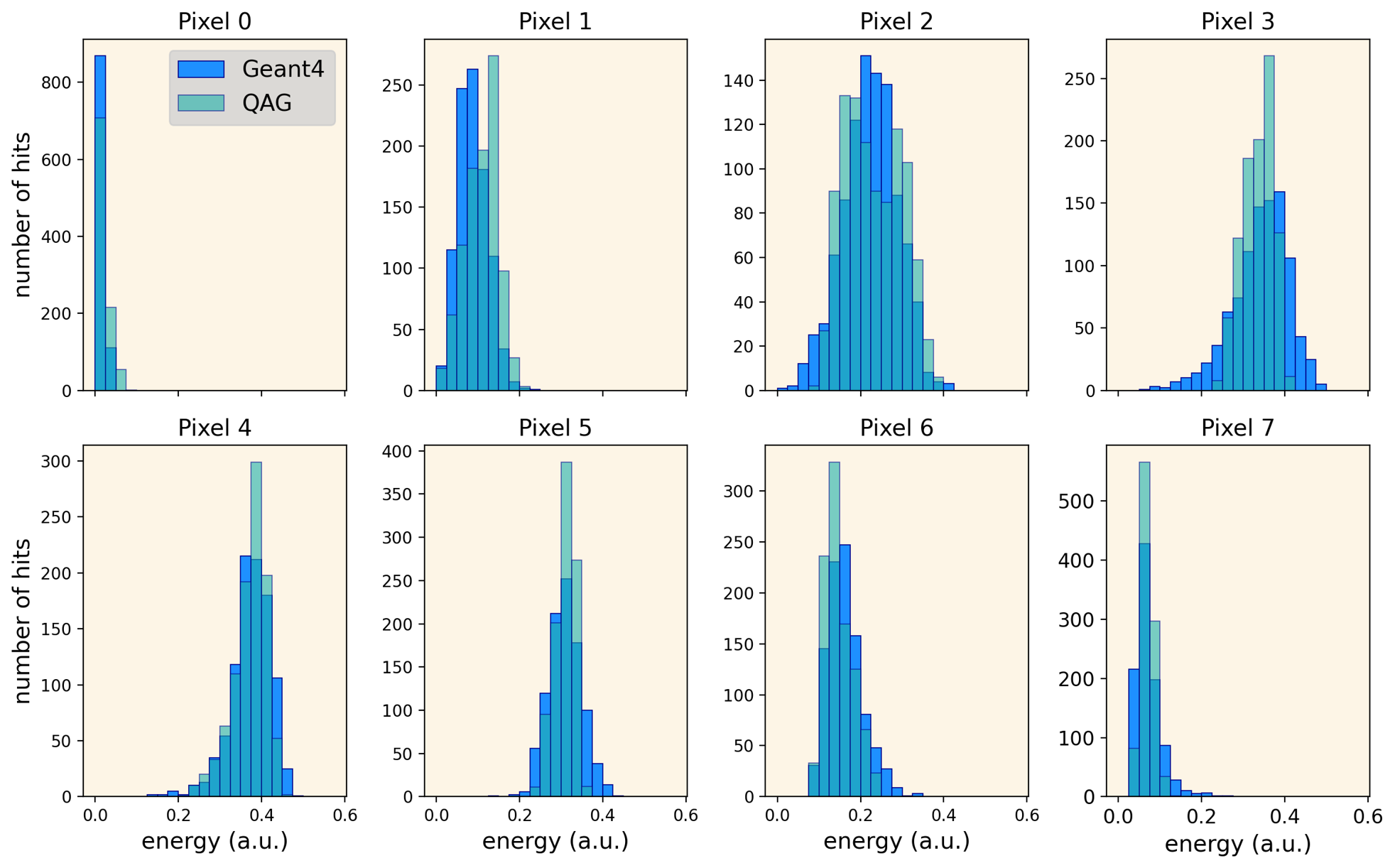}
    \caption{The energy deposition histograms for Geant4 and the QAG model. Each histogram represents the energy distribution for an individual pixel, as labeled above each plot. The x-axes represent the energies on a consistent scale across all histograms. The y-axes count the hits per energy bin, with each hit corresponding to the energy contained in a pixel for a single image. Note the varying y-axis scales.}
    \label{fig:pixel_wise_energy}
\end{figure}

\textbf{5. Pixel-wise energy distribution}: 
The last metric is used to examine  the distributions of the energy content of each pixel, as illustrated in figure \ref{fig:pixel_wise_energy}. Overall, the histograms of the QAG model match those of the Geant4 model. Even pixels with non-Gaussian energy distributions in the Geant4 model are correctly reproduced by the QAG model. For example, the longer tail towards smaller energies on the left side of pixel 4 is equally present in the histogram for the QAG model. The large histogram overlaps indicate that not only the averages are reproduced with high accuracy, but also the energy distributions returned for each individual pixel are correct.

\section{Quantum Noise Study}
In the current NISQ era, relatively high hardware error levels are one of the primary limitations to effectively employing algorithms on real quantum devices. Similar as in the classical case, QML models appear to be noise resilient to some degree of hardware errors \cite{noise_robust_circuits, QGAN_noise_study, QGAN_ACAT_noise_study, noise_robust_training}. In the following, the robustness of the QAG model to simulated noise is tested in inference and training. Furthermore, training and inference are executed on real quantum devices with measured noise levels and compared to the results with simulated noise.

\subsection{Inference}
In a first step, quantum noise is applied only to the inference of a model trained without noise. Inference is performed using three different noise configurations: simulated noise at varying levels, simulated noise derived from the later used hardware, and finally, with the real quantum hardware. In the simulated noise configurations, each qubit noise is modeled with the same readout measurement error, and each inter-qubit connection with the same two-gate (CNOT) error. In the combined noise model both are on the same level. In contrast, the hardware noise levels can vary widely for individual qubits as well as for gates.

\begin{figure}[t!]  
    \centering
    \includegraphics[width=0.95\textwidth, clip=true]{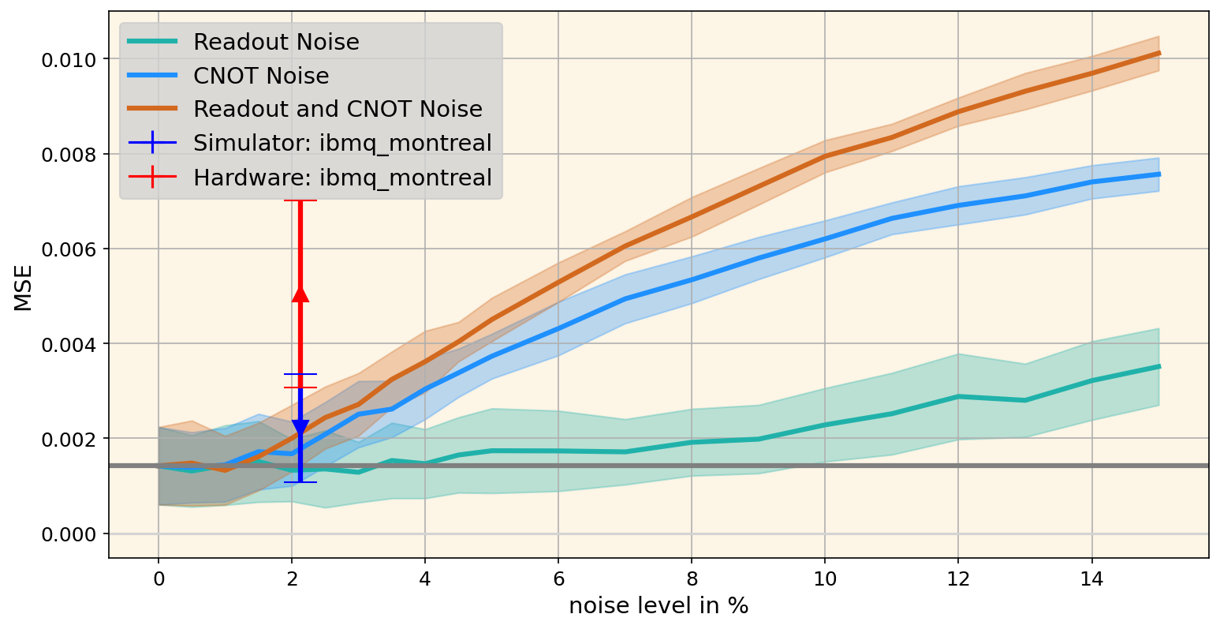}
    \caption{Inference noise study. On the x-axis the noise levels are provided and on the y-axis the MSE as accuracy metrics (the lower, the better). Inference is executed with various noise configurations and on real quantum hardware. At each point, twenty images are generated.}
    \label{fig:noisy_inference}
\end{figure}

Multiple noise configurations and error levels from zero up to 15\% are tested. The MSE is utilized as the accuracy measure. The results are illustrated in figure \ref{fig:noisy_inference}. For each configuration, the average value of 20 generated images is plotted in dependence of the noise level as a solid line and the standard deviation as a colored band around the mean. The gray horizontal line serves as accuracy reference for the noise-free configuration.
All configurations maintain sufficient accuracy up to approximately 1.5\% of noise. The configuration with readout noise only (green) is most robust and maintains a stable accuracy of up to 8\% of noise. CNOT noise only (blue) and both readout noise and CNOT noise combined (orange) experience a stronger impact. As expected, the combined configuration (orange) performs worst. As a side note, current quantum devices have average noise levels below 5\%, which is expected to gradually decrease further. But as discussed in the following, the noise levels are unstable and can sometimes spike up. Therefore, wider noise ranges were investigated.

Next, the inference is run by loading the real hardware noise model from \texttt{ibmq\_montreal} into the simulator. The \texttt{ibmq\_montreal} device consists of a Falcon r4 processor with 27 qubits. The average readout noise over the qubits employed at the time of the test is 2.51\% and the average gate noise level is 0.97\%. The explicit noise model, containing the entries for each qubit, is provided in the appendix in figure \ref{fig:noise_after_calib}. The result is included in figure \ref{fig:noisy_inference} (blue triangle). The noise level position (x-axis) is determined as the average of readout and CNOT noise. Although the noise levels of the qubits vary strongly, the measured accuracy of the hardware noise simulation agrees well with the simulated noise in mean and standard deviation within the uncertainties. This suggests that a model trained without any noise would theoretically be able to run inference on noisy hardware without a significant drop in accuracy. 

Finally, the inference is executed on the real \texttt{ibmq\_montreal} device. The result is plotted as a red triangle in figure \ref{fig:noisy_inference}. The accuracy on the real hardware is worse than predicted by the simulation, as indicated by a larger MSE value. The decomposition of the circuit to the real hardware includes swap operations, which imply additional two-qubit entanglement gates for the quantum circuit. It is possible that these are not included in the hardware noise simulation and lead to higher noise influence on real hardware and thus to worse results.

\subsection{Training}
The noise study is repeated to include noise also during training. We study if the QAG model can learn to compensate for noise in training, especially when running on the real quantum device. Besides, we investigate with which noise values the model can still maintain a reasonable accuracy.

\begin{figure}[t!]  
    \centering
    \includegraphics[width=0.98\textwidth, clip=true]{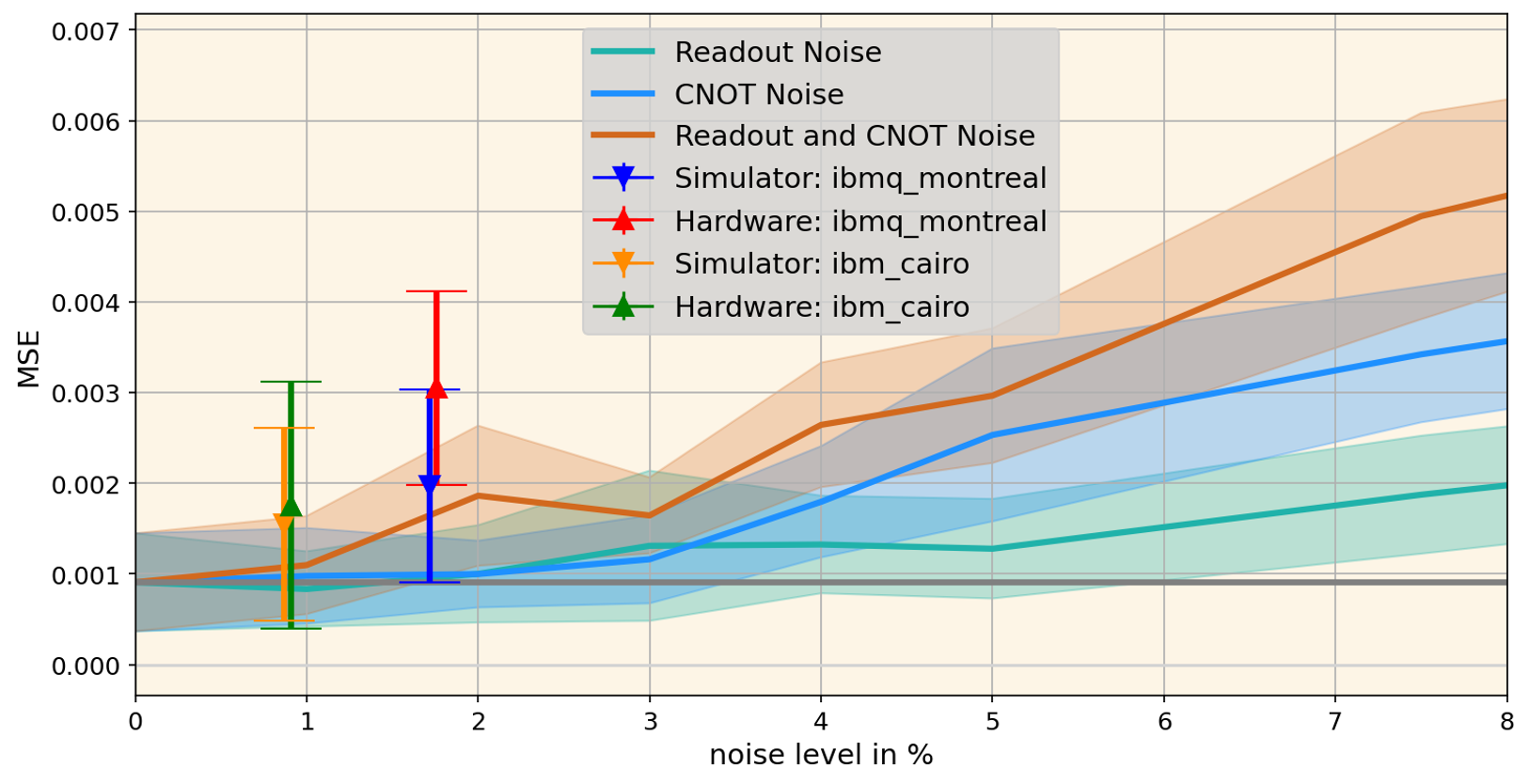}
    \caption{Training noise study. On the x-axis the noise levels are provided and on the y-axis the MSE as accuracy metrics (the lower, the better). Training is executed with various noise configurations and on real quantum hardware.
    The training with fake noise levels are repeated ten times.}
    \label{fig:noisy_training}
\end{figure}
The results are provided in figure \ref{fig:noisy_training}. The configurations with readout noise (green) and CNOT noise (blue) maintain a similar level of accuracy until approximately 3\%. The accuracy of the combination of readout and CNOT noise (orange) decreases marginally from 1\% of noise and further with larger noise levels. However, at 3\% of noise, its accuracy is still close to the noiseless case, staying within one standard deviation. This indicates that training the model with noise makes the QAG model more robust than only applying noise to a trained model in inference only.

Next, two quantum devices are simulated. \texttt{ibmq\_montreal} and \texttt{ibm\_cairo}, which are both 27 qubit devices, but \texttt{ibm\_cairo} has the more advanced Falcon r5.11 processor. Likewise, the training is repeated ten times and the results are added as blue and orange triangles in figure \ref{fig:noisy_training}. It can be noted that the average accuracy of the training with hardware noise performs slightly worse than the simulated combined noise model (orange line). The strong overlap between the error from the noise simulation (orange band) and the hardware noise simulation (orange and blue error bar) indicates that the accuracy difference is statistically not significant. 

\begin{figure}[t!]  
    \centering
    \begin{subfigure}[c]{.49\textwidth}
        \subcaption{}
        \includegraphics[width=\textwidth]{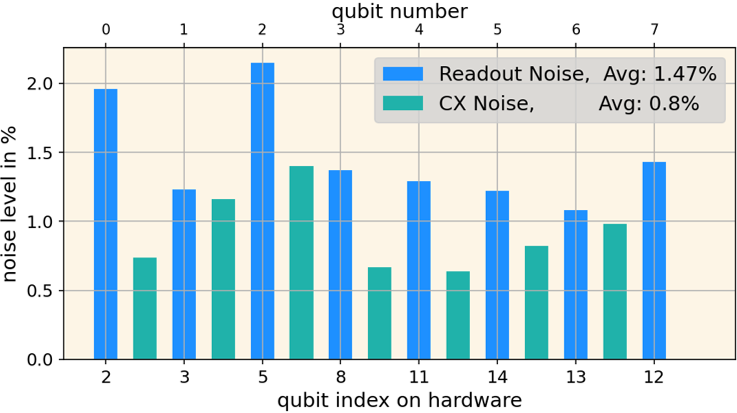}
        \label{fig:noise_before_calib}
    \end{subfigure}
    \hfill
    \begin{subfigure}[c]{.49\textwidth}
        \subcaption{}
        \includegraphics[width=\textwidth]{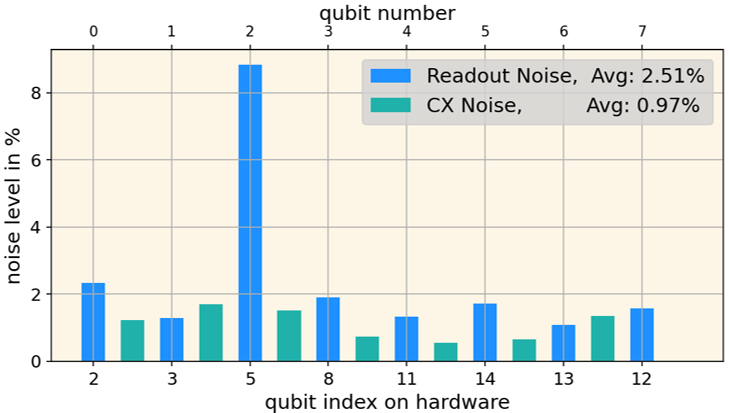}
        \label{fig:noise_after_calib}
    \end{subfigure}
    \caption{The noise levels of the \texttt{ibmq\_montreal} device are shown (a) before and (b) after the calibration change during training. The lower x-axes displays qubit indices as provided at the IBMQ hardware, while the upper x-axes represents their usage in the quantum circuit (e.g., in figure \ref{fig:circuit_architectures}, counting from top to bottom).
    Inference is performed with the noise model (b), where it is notable that the readout noise of qubit number five on the hardware increased from two to over eight percent. Note the different y-axis scales.}
   \label{fig:noise_levels}
\end{figure}

\begin{figure}[t!]  
    \centering
    \begin{subfigure}[c]{.48\textwidth}
        \subcaption{}
        \includegraphics[width=\textwidth]{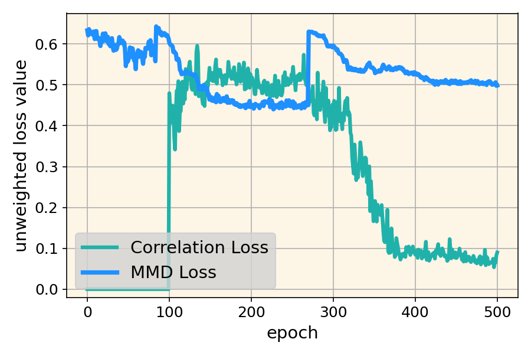}
        \label{fig:montreal_loss}
    \end{subfigure}
    \hfill
    \begin{subfigure}[c]{.48\textwidth}
        \subcaption{}
        \includegraphics[width=\textwidth]{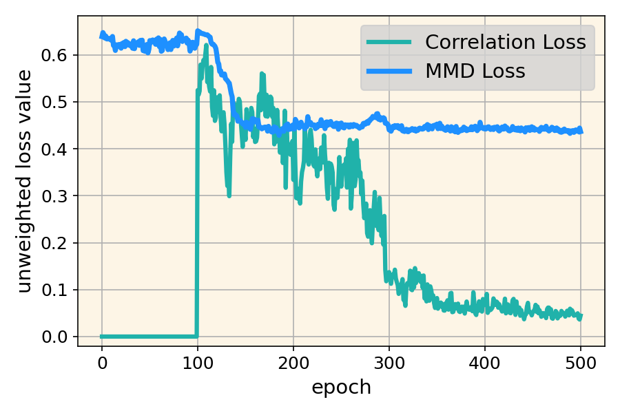}
        \label{fig:cairo_loss}
    \end{subfigure}
    \caption{The losses during training on the real quantum device for (a) \texttt{ibmq\_montreal} and (b) \texttt{ibm\_cairo}. Shown are the unweighted MMD and Correlation loss as a function of the training epochs.}
   \label{fig:hardware_losses}
\end{figure}

Finally, the entire training was executed on the real quantum device. First, the training was carried out on \texttt{ibmq\_montreal}. During training, around epoch 280, an unpredicted significant noise change occurred and the readout noise of one qubit increased to 8\%, as shown in figure \ref{fig:noise_levels}. As a result, the MMD loss spikes up, as shown in figure \ref{fig:montreal_loss}. This negatively influenced the training. However, after the noise change during training, the model recovered and adapted to the new noisy environment. In the still remaining number of epochs the loss decreased to a modest level. The training was repeated on the best performing \texttt{ibm\_cairo} device without having a hardware calibration change during training, and the training losses are shown in figure \ref{fig:cairo_loss}. 

The results of both hardware training are included as red and green triangles in figure \ref{fig:noisy_training}. The error bars correspond to the accuracy deviations within 50 generated validation images. It can be observed that the average accuracy of the \texttt{ibmq\_montreal} hardware training (red) is visibly worse than the noise simulation (blue). This is most likely due to the large noise increase and the fact that the model did not have enough remaining epochs to fully recover. This is also indicated by the still decreasing losses towards the end of the training. As mentioned above, the training was repeated on the more stable machine \texttt{ibm\_cairo}. This time, the training was completed without calibration changes and with about $1$\% lower hardware noise level (readout noise $0.86$\% and CNOT noise $0.89$\%). The accuracy of the hardware training on \texttt{ibm\_cairo} (green) is only marginally worse than the one from the simulator (orange). This indicates that the simulated and real hardware results behave similarly for low hardware noise levels. This fulfills the expectations derived from the pure simulation that exhibit only statistically insignificant variations in the very good accuracy at these low noise levels.

Comparing the absolute MSE magnitude of the real hardware training ($\approx0.002$ and $\approx0.003$) with the one from inference from the previous section ($\approx0.005$), the accuracy improved, suggesting that the QAG model is able to adapt its parameters to the noisy hardware to improve its precision. This is also confirmed by the \texttt{ibmq\_montreal} training, where the accuracy worsened entirely after the calibration change, but then recovered.  In the appendix in figure \ref{fig:montreal_shower_shape}, the average shower image created by the \texttt{ibmq\_montreal} device is visualized, and in figure \ref{fig:cairo_shower_shape} by \texttt{ibm\_cairo}. The shower image of \texttt{ibm\_cairo} agrees well with Geant4, whereas \texttt{ibmq\_montreal} exhibits some deficiencies because of the missing remaining training epochs after the calibration.

\section{Conclusion}
The results of this study clearly demonstrate that the newly developed QAG model is capable of generating images with good precision, as measured with a variety of validation metrics. This includes correctly reproducing average values, but most importantly, also complex pixel-wise correlations with the chosen optimal MERA-up quantum architecture. These results reveal that the QAG model with a good entangled circuit is capable of learning intrinsic correlation patterns from the training data.

Our study exhibit the significant impact that quantum hardware noise can have on the accuracy of quantum machine learning models. The results evidence that training the models with noise leads to better performance (stable until 3\% noise) because the QAG model adapts to the underlying noise behavior and converges faster in contrast to the situation of applying noise in inference only (stable until 1.5\% noise). This was also verified on the real hardware by the \texttt{ibm\_cairo} training. Furthermore, our study shows that the QAG model is robust and can produce accurate results even with significant hardware calibration changes with up to 8\% noise, as demonstrated by the training on \texttt{ibmq\_montreal}. Overall, the newly developed  QAG model demonstrates that training quantum machine learning models with realistic quantum hardware noise can lead to robust models and accurate results, which is of great importance for the future development of real world quantum machine learning applications.

\subsubsection*{Acknowledgment}
The authors would like to express their sincere gratitude to Simon Schnake and Alexis-Harilaos Verney-Provatas for their help, advice, and proofreading, which improved the quality and outcome of this paper.  
\\This work has been sponsored by the Wolfgang Gentner Programme of the German Federal Ministry of Education and Research and by the CERN Quantum Technology Initiative. Additionally, we would like to acknowledge partial funding from the BMBF Project "NiQ: Noise in Quantum Algorithms" within the BMBF program "Anwendungsnetzwerke für das Quantencomputing“ and from the Deutsches Elektronen-Synchrotron DESY, a member of the Helmholtz Association (HGF). 
\\Access to the IBM Quantum Services was obtained through the IBM Quantum Hub at CERN. The views expressed are those of the authors and do not reflect the official policy or position of IBM or the IBM~Q team.


\newpage
\begin{appendices}

\section{QAG Inference Algorithm}
\label{app:QAG_algorithm}
Algorithm~\ref{alg:algorithm} shows the pseudo code for generating one image using the QAG model from section~\ref{sec:QAG_description}. Here the MERA-up architecture (appendix~\ref{sec:A1}) is used as an example. 
The standard deviation $\sigma_{\text{pixel}}$ for the energy distribution of each pixel is obtained from training data.

\begin{algorithm}[h!]
\caption{Quantum Angle Generator Inference}\label{alg:algorithm}
\algblock[Name]{Begin}{End}
\algblockdefx[NAME]{START}{END}%
[2][Unknown]{Begin #1(#2)}%
{Ending}
\algblockdefx[NAME]{}{OTHEREND}%
[1]{Until (#1)}
\begin{algorithmic} 
    \State \textbf{Input}: $nb_{\textrm{shots}}$ \Comment{Number of measurements}\\
    \phantom{\textbf{Input}: }$\sigma_{\text{pixel}}$ \Comment{Array of the pixel standard deviations}
    \\
    \State \textbf{Output}: $E$ \Comment{Array of pixel energies forming the image}\\
    \rule{\linewidth}{0.4pt}

    \Begin
        \State Initialize all 8 qubits to state $\vert 0 \rangle$
        \State Randomly draw value $v$ from uniform distribution $[-0.25, 0.25]$
        \ForAll{index, qubits}
            \State Apply Hadamard gate H to qubit
            \State Randomly draw $\Omega$ from uniform distribution $[-1, 1]$
            \State $\Omega_{\text{pixel}}$ = $\Omega \cdot v \cdot \sigma_{\textrm{pixel}}[\textrm{index}]$
            \State Apply rotation Ry($\Omega_{\textrm{pixel}}$) to qubit
        \EndFor
        \State Apply the MERA-up unitary transformation to all qubits
        \State Measure the qubits $nb_{\textrm{shots}}$ times
        \State Count number of times state $\vert 0 \rangle$ is measured $\rightarrow \textrm{counts}(\vert 0 \rangle)$ 
        \State Compute $\theta$ according to equation~\ref{eq:Intersection}
        \State Compute pixel energies $E$ according to transformation~\ref{eq:decoding2}
        \State Return $E$
\End
\end{algorithmic}
\end{algorithm}

\section{Quantum Circuit Architectures under Study}\label{sec:A1}
\begin{figure}[ht!]  
    \centering
    \includegraphics[width=0.99\textwidth, clip=true]{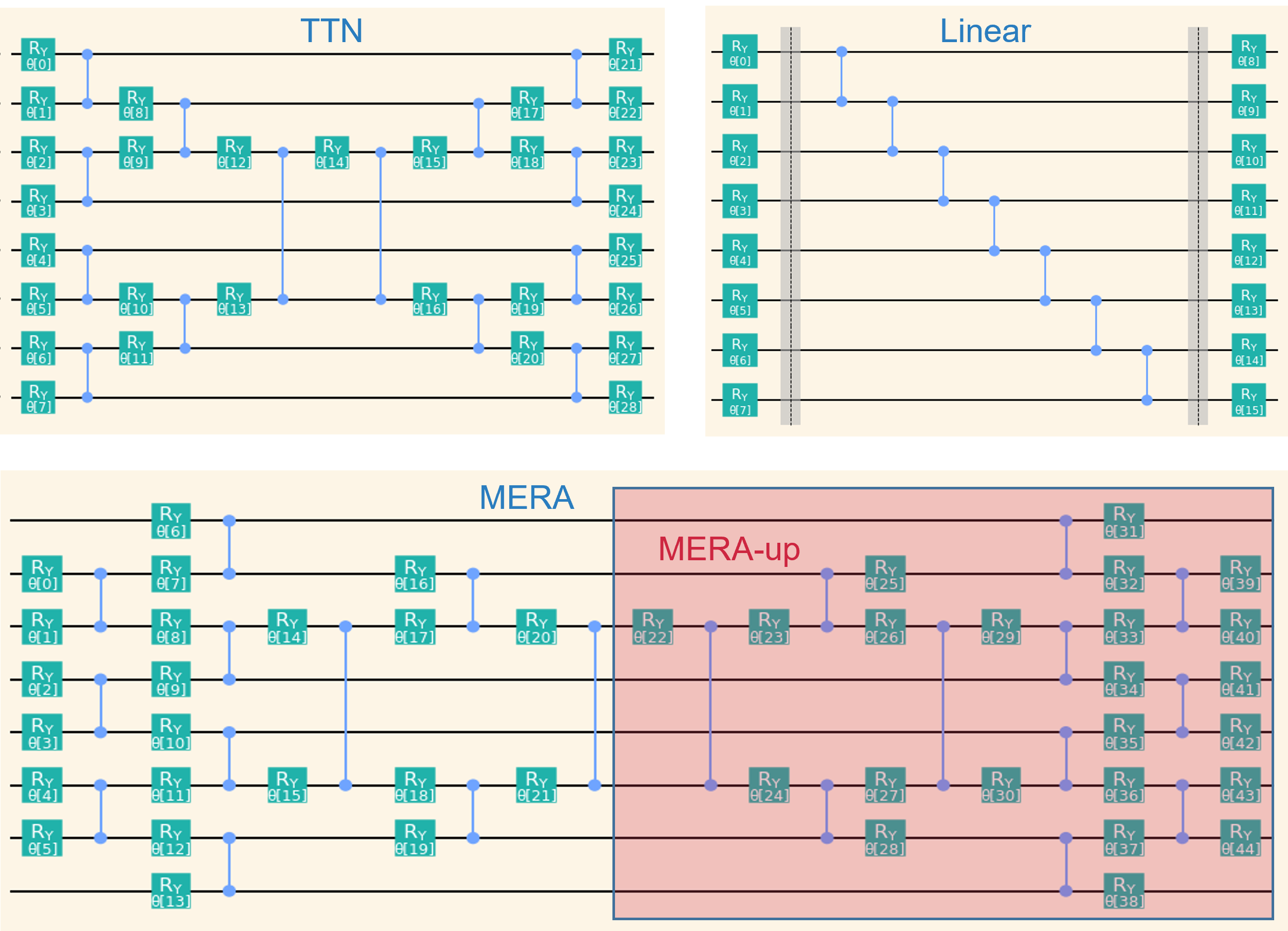}
    \caption{The basic quantum circuit architectures under study. Top left: TTN architecture, top right: Linear entanglement architecture, bottom: the full circuit corresponds to the MERA architecture, with the red highlighted section corresponding to the MERA-up architecture.}
    \label{fig:circuit_architectures}
\end{figure}

The quantum circuit architectures investigated in this paper are summarized in figure \ref{fig:circuit_architectures}. In general, it was observed that hierarchical architectures perform best while maintaining a reasonable number of quantum gates and parameters. Specifically, we examine the Tree Tensor Network (TTN) architecture and Multi-scale Entanglement Renormalization Ansatz (MERA) introduced in reference \cite{TTN_MERA}. Multiple variations of these circuits are tested: 1) circuits with a depth two (naming scheme "d2"). These circuits contain two layers, with all circuit gates placed twice, to evaluate if deeper circuits perform better. 2) Circuits additional with Rz-gates are employed after each Ry-gate (naming scheme "Rz") to assess if rotations around an additional axis can improve the accuracy. Both architecture variants double the number of parameters of the initial circuit.

For the MERA architecture, further variations are analyzed where only the right circuit half from the original architecture is implemented, denoted as the MERA upsampling (MERA-up) circuit. In the MERA-up circuit, the information is upsampled from the central qubit and spread to all the other qubits, similar to what happens in classical ML generative models, for example, transpose convolutional neural networks. The left half of the MERA circuit, the MERA downsampling (MERA-down) architecture, is not tested because it would rather compress the information as required in classification tasks that are not used in this paper. The last architecture contains a simple linear entanglement strategy.

In reference \cite{expressibility}, multiple complex four-qubit architectures are compared. However, these architectures are not exploited in our study. The reason is most of them contain many more gates and parameters, and they do not scale well for more than four qubit circuits. Also, many architectures employ parameterized two-qubit rotational gates which we found, taking the decomposition into account, not as effective as the combination of separate rotation and entanglement gates.

\section{Characteristic Circuit Numbers}\label{sec:A2}

\begin{figure}[ht!]  
    \centering
    \begin{subfigure}[c]{.325\textwidth}
        \subcaption{}
        \includegraphics[height=\textwidth]{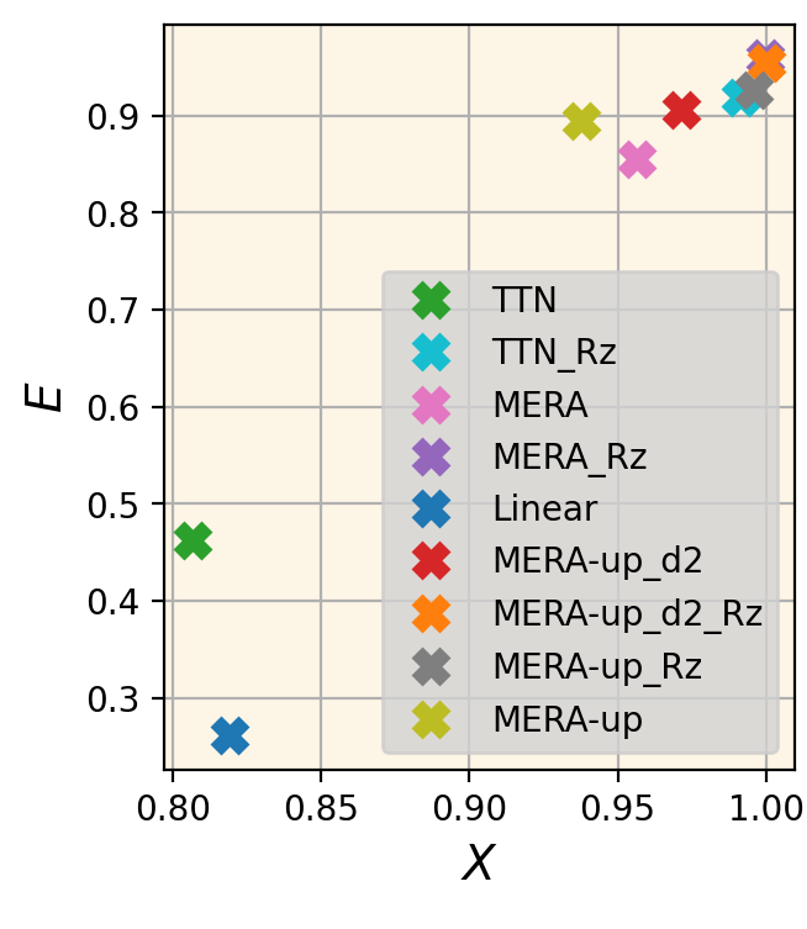}
        \label{fig:circuits_expr_ent}
    \end{subfigure}
    \begin{subfigure}[c]{.325\textwidth}
        \subcaption{}
        \includegraphics[height=\textwidth]{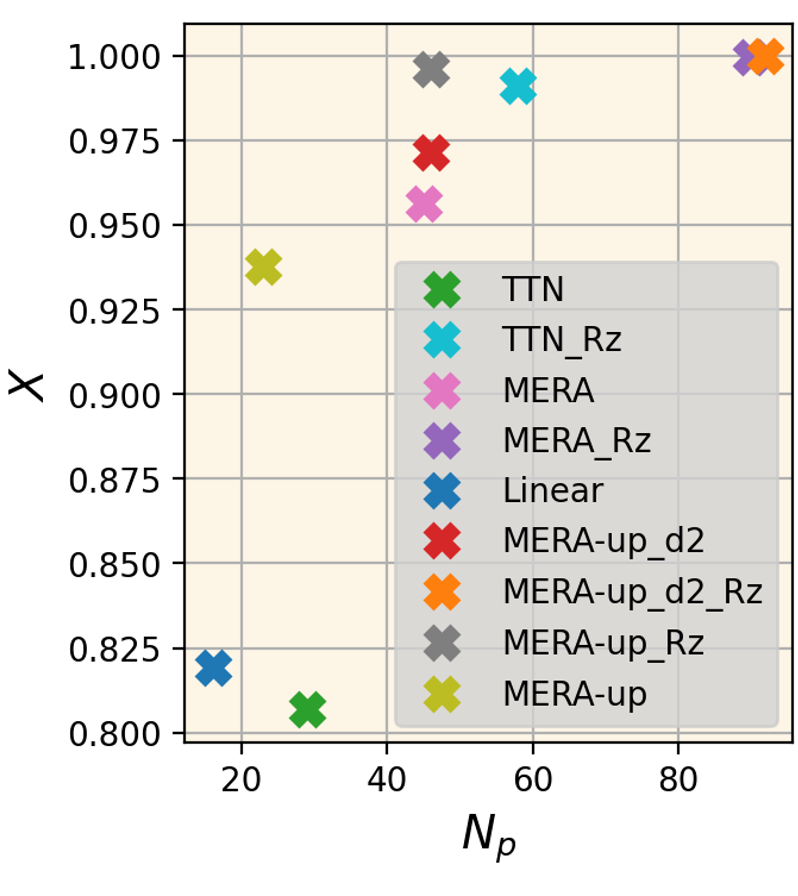}
        \label{fig:circuits_params_expr}
    \end{subfigure}
    \begin{subfigure}[c]{.325\textwidth}
        \subcaption{}
        \includegraphics[height=\textwidth]{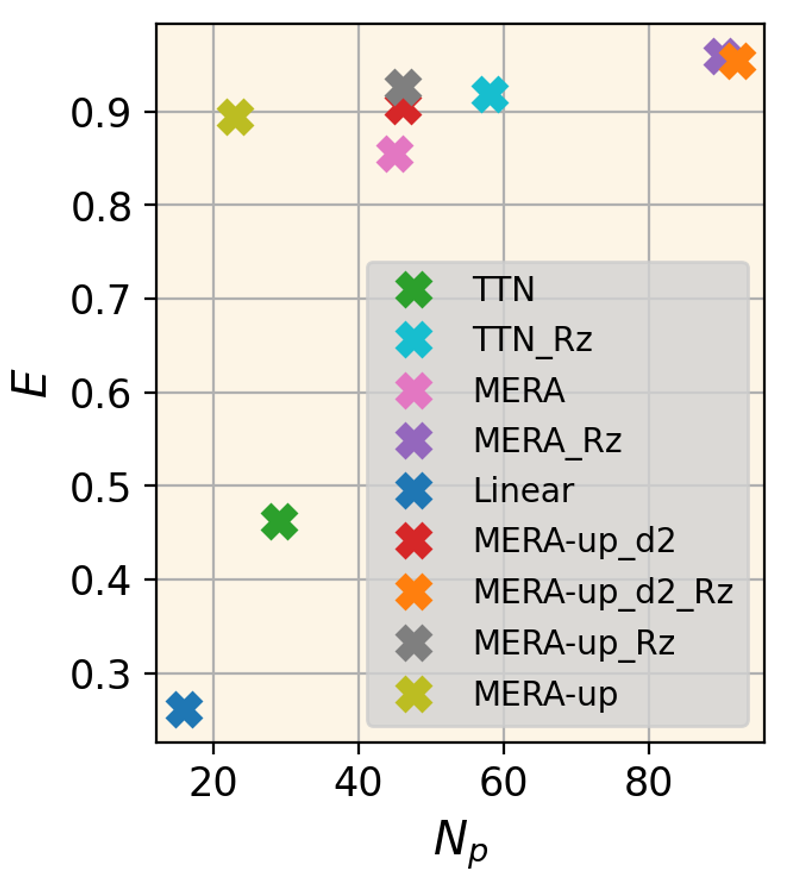}
        \label{fig:circuits_params_ent}
    \end{subfigure}
    \caption{Study of potential correlations between the three characteristic circuit numbers. On the left: $X$ versus $E$. In the middle: $N_p$ versus $X$. On the right: $N_p$ versus $E$.}
   \label{fig:circuits_characteristic_numbers}
\end{figure}

The results of evaluating the characteristic circuit numbers (number of parameters $N_p$, expressibility $X$ and entanglement capability $E$) are shown in figure \ref{fig:circuits_characteristic_numbers}. In figure \ref{fig:circuits_expr_ent} $E$ is plotted as a function of $X$. The MERA\_Rz and MERA-up\_d2\_Rz architecture perform best and lie almost on top of each other in the top right corner, directly followed by MERA-up\_Rz and TTN\_Rz. The Linear architecture and the basic TTN perform worst. It can be clearly noted that circuits with more gates (Rz, d2) perform better. $N_p$ is studied in figure \ref{fig:circuits_params_expr} and figure \ref{fig:circuits_params_ent}. It can be seen that the best two circuits -- MERA\_Rz and MERA-up\_d2\_Rz -- contain by far the largest number of parameters. However, a more limited number of parameters is desirable to address NISQ hardware limitations. The MERA-up\_Rz and TTN\_Rz architectures contain approximately half the number of parameters compared to the MERA\_Rz and MERA-up\_d2\_Rz circuits but perform almost similarly accurately. Therefore, they may be preferred in practice. Further reducing the number of parameters, the MERA-up architecture with only $N_p=23$ parameters maintains adequate values for $E$ and $X$ and is the baseline circuit for the more detailed studies. All characteristic circuit numbers measured in the circuit study are summarized in table \ref{tab:circuits}. In the table, it is evident that the MERA-up circuit maintains a considerably lower training time on the classical hardware compared to models with a higher number of parameters. The training was executed on a cluster equipped with 32 Intel Xeon Gold 6130 CPU cores running at 2.10GHz.

\begin{table}[t]
\center
\caption{The characteristic circuit values $N_p$, $X$ and $E$ are listed. Additionally, the statistics of the MSE accuracy metrics and the training times $t$ in minutes are provided for the individual circuits for repeating the training 25 times on the calorimeter data set.}
\begin{tabular}{c | c | c | c | c | c}
 Circuit Name           & $N_p$ &  $X$ & $E$ & MSE & $t$ in min \\
 \hline
 Linear                 &  16        &  0.8191    &     0.261                           & 0.00113 $\pm$ 0.0008  & 35.0 $\pm$ 0.2 \\[0.1cm]
 TTN                    &  29        &  0.8068    &     0.462                           & 0.00245 $\pm$ 0.0012   &  41.5 $\pm$ 0.6 \\ [0.1cm]
 TTN\_Rz                 &  58        &  0.9912    &     0.918                           & 0.00440 $\pm$ 0.0022  &  85.9 $\pm$ 1.5  \\[0.1cm] 
 MERA                   &  45        &  0.9564    &     0.855                           & 0.00220 $\pm$ 0.0011  &  51.3 $\pm$ 0.5  \\[0.1cm] 
 MERA\_Rz                &  90        &  0.9997    &     0.959                           & 0.00634 $\pm$ 0.0036 &  121.5 $\pm$ 9.6  \\[0.1cm] 
 \textbf{MERA-up}                 &  \textbf{23}        &  \textbf{0.9377}    &     \textbf{0.894}                           & \textbf{0.00059} $\pm$ \textbf{0.0004}  &  39.4 $\pm$ 2.9   \\[0.1cm] 
 MERA-up\_d2             &  46        &  0.9715    &     0.906                           & 0.00038 $\pm$ 0.0002  &  50.5 $\pm$ 0.2   \\[0.1cm] 
 MERA-up\_Rz             &  46        &  0.9961    &     0.926                           & 0.00047 $\pm$ 0.0005  &  75.2 $\pm$ 7.5   \\[0.1cm] 
 MERA-up\_d2\_Rz          &  92        &  0.9999   &      0.954                           & \;0.00094 $\pm$ 0.0008  &  116.2 $\pm$ 4.3   
\label{tab:circuits}
\end{tabular}
\end{table}

\section{Definitions of Accuracy Metrics }
\label{app:validation_code}
In this section, we provide detailed formulas for the inference accuracy metrics discussed in section \ref{sec:accuracy_evaluation}.

\subsection{MSE of the Average Shower Shape}
The Mean Squared Error (MSE) for the average calorimeter shower shape is derived from the difference between the average images of Geant4 and the QAG model, calculated over 980 generated inference images and the corresponding 980 Geant4 images. In the subsequent equations, $k$ iterates over the $N=$ 980 images and $i$ denotes the i\textsuperscript{th} pixel of each image. Hence, $E_{k, \, i}^\textrm{G4}$ (Geant4) and $E_{k, \, i}^\textrm{QAG}$ (QAG model) correspond to the energy of pixel $i$ from image $k$. The average pixel energies $\overline{E}_i^{\textrm{G4}}$ and $\overline{E}_i^{\textrm{QAG}}$ are computed as follows:
\begin{equation}
    \overline{E}_{i}^\textrm{G4} = \frac{1}{N} \sum_{k=1}^{N} E_{k,  \, i}^\textrm{G4} 
    \hspace{0.5cm} \text{and} \hspace{0.5cm}
    \overline{E}_{i}^\textrm{QAG} = \frac{1}{N} \sum_{k=1}^{N} E_{k,  \, i}^\textrm{QAG}   .
\end{equation}

The MSE is then calculated by iterating over the $M=$ 8 individual pixels by the index $i$:
\begin{equation}
\text{MSE} = \frac{1}{M} \sum_{i=1}^{M} \Big( \overline{E}_i^\textrm{G4} - \overline{E}_i^\textrm{QAG} \Big)^2  .
\end{equation}
As the MSE corresponds to an error, the smaller the value, the better the accuracy of the QAG model.


\subsection{Energy Sum Histogram}
Each histogram entry (or hit) is the total energy $\hat{E}$ of a shower image: 
\begin{equation}
\hat{E}^\textrm{G4} = \sum_{i=1}^{M} E_{i}^\textrm{G4} 
\hspace{0.5cm} \text{and} \hspace{0.5cm}
\hat{E}^\textrm{QAG} = \sum_{i=1}^{M} E_{i}^\textrm{QAG}  ,
\end{equation}
where $i$ is iterating over the $M=8$ pixels.

The means of the distributions for our data sets, Geant4 and QAG, are denoted as $\mu^\textrm{G4}$ and $\mu^\textrm{QAG}$, respectively. Similarly, the standard deviations are represented by $\sigma^\textrm{G4}$ and $\sigma^\textrm{QAG}$. 

The means $\mu$ of the total energy for the $N=980$ images are then calculated using the well-known equation:
\begin{equation}
\mu^\textrm{G4} = \frac{1}{N} \sum_{i=1}^{N} \hat{E}_{\textrm{i}}^\textrm{G4} 
\hspace{0.5cm} \text{and} \hspace{0.5cm}
\mu^\textrm{QAG} = \frac{1}{N} \sum_{i=1}^{N} \hat{E}_{\textrm{i}}^\textrm{QAG}  . 
\end{equation}

And the standard deviations, denoted by $\sigma$, are computed as:
\begin{equation}
\begin{split}
\sigma^\textrm{G4} &= \sqrt{\frac{1}{N} \sum_{i=1}^{N} (\hat{E}_{\textrm{i}}^\textrm{G4} - \mu^\textrm{G4})^2} \\ \\
&\text{and} \\ \\
\sigma^\textrm{QAG} &= \sqrt{\frac{1}{N} \sum_{i=1}^{N} (\hat{E}_{\textrm{i}}^\textrm{QAG} - \mu^\textrm{QAG})^2}
\end{split}
\end{equation}

A closer agreement between the means and standard deviations of the two distributions indicates a higher accuracy of the QAG model.

\section{Full Training on Quantum Device}\label{sec:hardware_training}
\begin{figure}[ht!]  
    \begin{subfigure}[c]{.26\textwidth}
        \subcaption{}
        \includegraphics[height=\textwidth]{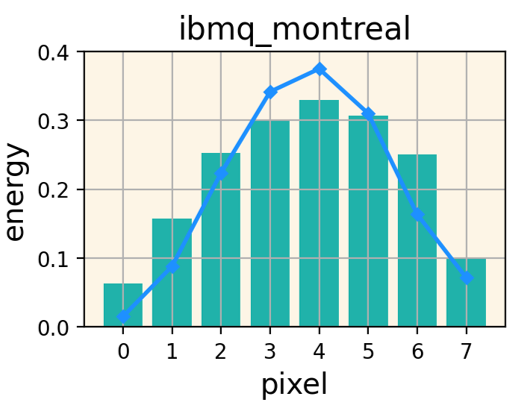}
        \label{fig:montreal_shower_shape}
    \end{subfigure}
    \hspace{0.7cm}
    \begin{subfigure}[c]{.26\textwidth}
        \subcaption{}
        \includegraphics[height=\textwidth]{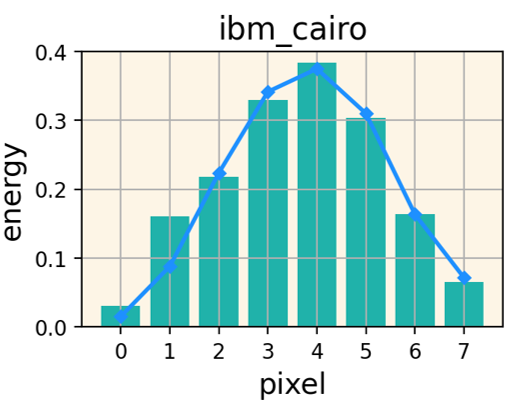}
        \label{fig:cairo_shower_shape}
    \end{subfigure}
    \hspace{0.7cm}
    \begin{subfigure}[c]{.25\textwidth}
        \subcaption{}
        \includegraphics[height=\textwidth]{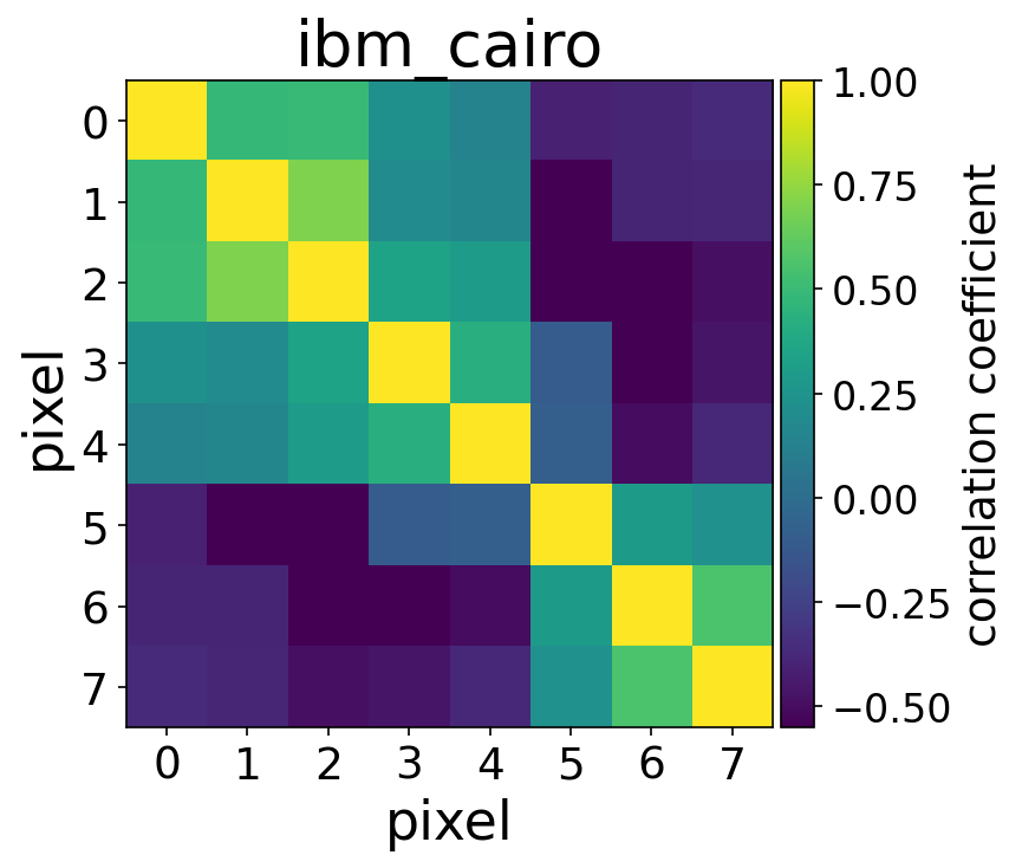}
        \label{fig:cairo_corr}
    \end{subfigure}
    \caption{On the left, the average shower images of the QAG model are trained on \texttt{ibmq\_montreal} and in the middle on \texttt{ibm\_cairo}. The correlation plot on the right is created for the \texttt{ibm\_cairo} training.}
    \label{fig:shower_images}
\end{figure}
The average shower images of the quantum hardware training are correctly reproduced, as provided in figure \ref{fig:shower_images}. The correlation plot for the \texttt{ibm\_cairo} training is provided in figure \ref{fig:cairo_corr}. The overall correlation pattern is correctly reproduced. All in all, the model trained on \texttt{ibm\_cairo} shows a good performance.

\end{appendices}

\newpage
\bibliography{bib}


\end{document}